\documentclass[onefignum,onetabnum]{siamart190516}
 
\usepackage{lipsum}
\usepackage{amsfonts}
\usepackage{graphicx}
\usepackage{epstopdf}

\newsiamremark{remark}{Remark}
\newsiamremark{hypothesis}{Hypothesis}
\crefname{hypothesis}{Hypothesis}{Hypotheses}
\newsiamthm{claim}{Claim}

\usepackage{amsopn}
\usepackage{hyperref}       
\usepackage{url}            
\usepackage{booktabs}       
\usepackage{amsfonts}       
\usepackage{nicefrac}       
\usepackage{microtype}      
\usepackage{xcolor}         
\usepackage{amsmath}
\usepackage{amssymb}
\usepackage{graphicx}
\usepackage{tabularx}
\usepackage{makecell}
\usepackage{listings}
\usepackage{paralist}
\usepackage{color}
\usepackage{siunitx}
\usepackage{multirow}
\usepackage{bm}
\usepackage{algorithm}
\usepackage{algpseudocode}
\usepackage{movie15}
\usepackage{siunitx}

\DeclareMathOperator{\diag}{diag}

\DeclareSIUnit{\cal}{cal}
\DeclareSIUnit{\kcal}{\kilo\cal}

\newcommand{\LB}[1]{#1}
\newcommand{\LBB}[1]{#1}
\newcommand{\Ldyn}{L^{dyn}}
\newcommand{\Lcondt}{L^{con}}

\newcommand{\Rd}{\mathbb{R}^d}
\newcommand{\R}{\mathbb{R}}
\newcommand{\secondterm}{\vect{g}}
\newcommand{\vect}[1]{\mathbf{#1}}
\newcommand{\mat}[1]{#1}

\ifpdf
\hypersetup{
  pdftitle={Computing the Invariant Distribution of Randomly Perturbed Dynamical Systems Using Deep Learning},
  pdfauthor={B. Lin, Q. Li, and W. Ren}
}
\fi

\headers{Computing Invariant Distributions Using Deep Learning}{B. Lin, Q. Li, and W. Ren}
\title{Computing the Invariant Distribution of Randomly Perturbed Dynamical Systems Using Deep Learning}

\author{Bo Lin\thanks{Department of Mathematics, National University of Singapore, Singapore 119076
  (\email{matboln@nus.edu.sg}, \email{qianxiao@nus.edu.sg}, \email{matrw@nus.edu.sg}).}
\and Qianxiao Li\footnotemark[1]
\and Weiqing Ren\footnotemark[1]}

\begin{document}

\maketitle

\begin{abstract}
The invariant distribution, which is characterized by the stationary Fokker-Planck equation, is an important object in the study of randomly perturbed dynamical systems. 
Traditional numerical methods for computing the invariant distribution based on the Fokker-Planck equation, such as finite difference or finite element methods, are limited to low-dimensional systems due to the curse of dimensionality. In this work, we propose a deep learning based method to compute the generalized potential, {\it i.e.} the negative logarithm of the invariant distribution multiplied by the noise.
The idea of the method is to learn a decomposition of the force field, as specified by the Fokker-Planck equation, from the trajectory data. 
The potential component of the decomposition gives the generalized potential.
The method can deal with high-dimensional systems, possibly with partially known dynamics.
Using the generalized potential also allows us to deal with systems at low temperatures, where the invariant distribution becomes singular around the metastable states. 
These advantages make it an efficient method to analyze invariant distributions for practical dynamical systems. 
The effectiveness of the proposed method is demonstrated by numerical examples.
\end{abstract}



\section{Introduction}
The probability density function of a randomly perturbed dynamical system is of great importance in studying its steady-state properties and transition events\LB{~\cite{wang2008potential,li2013quantifying,li2018landscape}}. The long-term effects of the noise on the dynamics can be investigated through the invariant distribution of the system, \LBB{for example, in biological networks~\cite{li2018landscape} and the socio-economic systems~\cite{furioli2017fokker}}.
In particular, a theoretical framework based on the underlying potential landscape, which is derived from the invariant distribution, can be used to analyze the robustness and stability of nonequilibrium systems~\cite{wang2008potential}.

Despite its analytical usefulness, numerical computation of the invariant distribution remains a central challenge for high dimensional systems, especially at low temperatures.
The invariant distribution is governed by the stationary Fokker-Planck equation.
Traditional numerical methods such as the finite difference method~\cite{sepehrian2015numerical}, the finite element method~\cite{galan2007stochastic} and the variational iteration  method \cite{torvattanabun2011numerical} have been used to effectively solve the Fokker-Planck equation in low dimensions. These methods require the discretization of a bounded domain in space, thus the computational cost usually increases exponentially with the dimension of the problem, a difficulty known as the curse of dimensionality. For this reason, these traditional numerical methods become prohibitively expensive  for practical systems where the dimension is larger than three\LB{~\cite{li2018landscape,chen2018efficient,li2013quantifying}}.
An alternative approach for computing the probability distribution is the Monte Carlo method. For example, in Ref.~\cite{zhai2020deep}, the probability density is estimated using the direct Monte Carlo method by sampling long trajectories of the stochastic differential equation or using the conditional Gaussian framework\LB{~\cite{chen2017beating,chen2018efficient}}.  A naive application of the Monte Carlo method suffers from the difficulty caused by meta-stability in systems with multiple meta-stable states, especially when the amplitude of the noise, {\it i.e.} the temperature is low. 

Recently, a number of deep learning based methods haven been proposed for solving partial differential equations (PDEs)~\cite{han2018solving,weinan2018deep,li2019computing,khoo2019solving,nabian2018deep,raissi2019physics,zang2020weak}. These methods have been very successful even for problems in high dimensions. 
In the work of Ref.~\cite{han2018solving}, a deep learning framework was designed for solving semilinear parabolic PDEs based on a reformulation of backward stochastic differential equations. In Refs.~\cite{weinan2018deep,li2019computing,khoo2019solving,nabian2018deep}, the solution of PDEs is approximated by neural networks and computed by solving the corresponding variational problems. 
In Ref.~\cite{raissi2019physics}, physics-informed neural networks (PINN) were introduced to compute solutions of PDEs.

In this paper, we focus on the stationary Fokker-Planck equation and develop a deep learning based method for computing the invariant distribution of randomly perturbed dynamical systems modeled by stochastic differential equations. Instead of computing the invariant distribution directly, we propose to compute the generalized potential, which is the negative logarithm of the invariant distribution multiplied by the noise. 
The method is based on a decomposition of the force field as specified by the Fokker-Planck equation. The potential component of the decomposition gives the generalized potential. 
We design the loss functions to learn the decomposition for both known and unknown force fields. In the latter case, the decomposition is learned from trajectory data of the corresponding deterministic dynamics. 
Thus the method is applicable for high-dimensional systems with partially known information of the dynamics at low-temperature regimes. 
The ability of the proposed method to compute the invariant distribution of practical systems at various temperatures and of high dimensions is demonstrated in model systems.

\LBB{The idea of learning a decomposition of the force field was used to compute the quasipotential~\cite{lin2021data}. The quasipotential describes the asymptotic property of the dynamics and characterizes the generalized potential in the zero noise limit~\cite{freidlin2012random,zhou2016construction,lin2021data}. The quasipotential satisfies a first-order Hamilton-Jacobi (HJ) equation. In Ref.~\cite{lin2021data}, the HJ equation was solved by learning an orthogonal decomposition of the force field from the trajectory data. In the current work, we use the similar idea to solve the Fokker-Planck equation for the invariant distribution at finite noise.}


The proposed method has advantages over the recently proposed deep learning based methods for solving the Fokker-Planck equation~\cite{xu2020solving,zhai2020deep,chen2021solving}. The main idea of the previous methods was to use a neural network to represent the probability density function and \LBB{then minimize a loss function involving the residual of the Fokker-Planck equation.}  
These methods become less efficient when the magnitude of the noise is low, as the density function becomes singular around the metastable states. 
In contrast, parameterizing the generalized potential in the current method allows us to deal with systems at low temperatures. Furthermore, the proposed method is data-driven in the sense that it can deal with systems with partially known dynamics. 

The paper is organized as follows. We first review the Fokker-Planck equation and some earlier work for solving the equation in Section~\ref{Realted_work}. In Section~\ref{Methods}, we propose the deep learning based method, including the decomposition of the force field, its parameterization using neural networks and the loss functions under two different problem settings.
The effectiveness of the method is demonstrated by numerical examples in Section~\ref{C5:Numerical_Examples}. We draw conclusions in Section~\ref{C5:Conclusion}.

\section{\LB{The Fokker-Planck equation and related work}}~\label{Realted_work}
Consider a dynamical system in $\Rd$ modeled by the stochastic differential equation (SDE)
\begin{equation}\label{C5:SDE}
d\vect{x}_t = \vect{f}(\vect{x}_t) dt+\sqrt{2\epsilon}\mat{\sigma} d\mathbf{W}_t,\quad t>0
\end{equation}
where $\vect{f}(\vect{x})$ is a vector (force) field, $\mathbf{W}_t$ is a $m$-dimensional Wiener process, $\mat{\sigma}\in \R^{d\times m}$ is a constant matrix, and $\epsilon>0$ is a parameter controlling the strength of the noise.
The invariant probability density function of the dynamical system, $p(\vect{x})$,
solves the Fokker-Planck (FP) equation
\begin{equation}\label{C5:FP}
  \mathcal{N} p(\vect{x}) :=  -\nabla\cdot (\vect{f}(\vect{x})p(\vect{x})) + \epsilon\nabla\cdot(D\nabla p(\vect{x}))=0,\quad \vect{x} \in \Rd
\end{equation}
where $\epsilon D=\epsilon\mat{\sigma}\mat{\sigma}^T\in \R^{d\times d}$ is the diffusion tensor, $\nabla p(\vect{x})$ and $\nabla\cdot \vect{q}(\vect{x})$ denote the gradient of function $p(\vect{x})$ and the divergence of the vector field $\vect{q}(\vect{x})$, respectively. The FP equation can also be written as $\nabla\cdot \vect{J}(\vect{x})=0$, where the probability flux $\vect{J}(\vect{x})=-\vect{f}(\vect{x})p(\vect{x}) + \epsilon D\nabla p(\vect{x})$. 

In low dimensions when $d\le 3$, the invariant distribution of the dynamical system \eqref{C5:SDE} can be computed by solving the FP equation using traditional numerical methods, \textit{e.g.} finite difference or finite element methods\LB{~\cite{sepehrian2015numerical,galan2007stochastic}}. Take the finite difference method for example. The FP equation is restricted to a bounded domain $\Omega$ with certain boundary condition, \textit{e.g.} the no-flux condition $\mathbf{J}(\vect{x})\cdot\mathbf{n}(\vect{x})=0$, $\vect{x}\in\partial\Omega$. where $\mathbf{n}$ is the outward normal vector on the boundary $\partial\Omega$. Then the differential operators in the equation are approximated by difference operators on a mesh covering $\Omega$, and the resulting difference equations are solved for the density function on the grid points of the mesh. Such mesh-based numerical methods can only be applied to low-dimensional systems. For practical systems when the dimension is greater than three, these methods become too expensive as the computational cost increases exponentially with the dimension. 

Recently, learning methods based on artificial neural networks were proposed to compute invariant distributions. 
In Ref.~\cite{xu2020solving}, it was proposed to solve the FP equation by parameterizing the solution using a neural network $p_{\theta}$. 
The neural network weights are trained by minimizing the loss function
\begin{equation}\label{loss_r1}
\begin{aligned}
    L      &= \int_{\Omega} \lvert \mathcal{N} p_{\theta}(\vect{x})\rvert^2 d\vect{x} + 
    \lambda_1 \left\lvert \int_{\Omega} p_{\theta}(\vect{x}) d\vect{x}-1\right\rvert^2 +\lambda_2 \int_{\partial \Omega} \lvert p_{\theta}(\vect{x})\rvert^2 d\vect{x},
\end{aligned}
\end{equation}
where the last two terms are introduced to impose the normalization condition and the homogeneous Dirichlet boundary condition respectively, $\lambda_1$, $\lambda_2$ are parameters controlling the proportion of the two penalty terms.
In Ref.~\cite{xu2020solving}, the integral in the second term was approximated using quadrature on a uniform mesh covering $\Omega$. 
This limited the applicability of the method to low-dimensional systems.

In another learning-based method~\cite{zhai2020deep}, the invariant density function was also parameterized by a neural network $p_\theta$. The neural network weights were trained by minimizing the loss function 
\begin{equation}\label{loss_r2}
    L = \int_{\Omega} \lvert \mathcal{N} p_{\theta}(\vect{x})\rvert^2 d\mu_X(\vect{x}) + 
    \int_{\Omega} \left\lvert p_{\theta}(\vect{y})-\tilde{p}(\vect{y}) \right\rvert^2
    d\mu_Y(\vect{y}),
\end{equation}
where $\mu_X(\vect{x})$ and $\mu_Y(\vect{y})$ are probability measures, $\tilde{p}(\vect{y})$ is a rough estimate of the invariant probability density obtained by sampling trajectories of the SDE~\eqref{C5:SDE}. To emphasize high probability regions, $\mu_X(\vect{x})$ and $\mu_Y(\vect{y})$ were chosen based on the sampled trajectories of the SDE~\eqref{C5:SDE} in Ref.~\cite{zhai2020deep}.

Both learning-based methods solve the FP equation for invariant density function $p(\vect{x})$ directly. When the noise $\epsilon$ is small, the density function becomes rather singular with peaks at meta-stable states and nearly zero elsewhere. In this situation, directly computing the density function becomes less efficient and may lead to inaccurate solutions. 

\section{Computing the generalized potential}\label{Methods}

Let $V(\vect{x})=-\epsilon\log p(\vect{x})$, where $p(\vect{x})$ is the invariant probability density function of the dynamical system \eqref{C5:SDE}. The function $V(\vect{x})$ is called the generalized potential. 
In a gradient system with the force field $\vect{f}(\vect{x})=-\nabla U_p(\vect{x})$ \LBB{and the diffusion matrix $D=I_d$}, where $U_p(\vect{x})$ is the potential function \LBB{and $I_d$ denotes the $d$-dimensional identity matrix}, the invariant distribution is given by the Boltzmann-Gibbs distribution: $p(\vect{x})=Z^{-1}\cdot e^{-\epsilon^{-1}U_p(\vect{x})}$, where $Z$ is the normalization constant. In this case, the generalized potential differs from the potential function by a constant: $V(\vect{x})=U_p(\vect{x})+c$, for all $\vect{x}\in\Rd$. 


\LBB{{\bf Remark.} {\it In general dynamical systems, the generalized potential is related to the global quasipotential $U(\vect{x})$, which can be constructed from local quasipotentials~\cite{freidlin2012random,zhou2016construction}. The local quasipotential is defined as the minimum action to reach $\vect{x}$ from a metastable state~\cite{freidlin2012random}. The global quasipotential characterizes the invariant distribution of the dynamical system in the zero noise limit~\cite{zhou2016construction,lin2021data} up to a constant: $\lim_{\epsilon\rightarrow0}\epsilon\log p(\vect{x})=-U(\vect{x})+c$. Therefore, the generalized potential $V(\vect{x})$ converges to the quasipotential $U(\vect{x})$ in the zero noise limit.}}

Using the ansatz $p(\vect{x})=e^{-\epsilon^{-1}V(\vect{x})}$ in the FP equation, we immediately obtain the following equation for the generalized potential:
\begin{equation}\label{FP2}
\nabla V(\vect{x})^T (\vect{f}(\vect{x})+D\nabla V(\vect{x})) - \epsilon \nabla\cdot (\vect{f}(\vect{x})+ D\nabla V(\vect{x})) = 0,\quad \vect{x}\in \Rd
\end{equation}
where we have dropped the exponential factor $e^{-\epsilon^{-1}V(\vect{x})}$.
Let $\secondterm(\vect{x})=\vect{f}(\vect{x})+ D\nabla V(\vect{x})$. Then solving the above equation is equivalent to finding a decomposition of the force field
\begin{equation}\label{C5:form}
    \vect{f}(\vect{x}) = -D\nabla V(\vect{x})+\secondterm(\vect{x}),
\end{equation}
such that 
\begin{equation}\label{C5:form2}
    \nabla V(\vect{x})^T \secondterm(\vect{x}) - \epsilon \nabla\cdot \secondterm(\vect{x})=0.
\end{equation}
For convenience, we call the first term in the decomposition~\eqref{C5:form} \LB{as} the potential component of the force field and the term $\secondterm(\vect{x})$ is referred to as the residual component.
Once the decomposition is found, we readily obtain the invariant distribution: 
$p(\vect{x})=e^{-\epsilon^{-1}V(\vect{x})}$ \LB{, where $V$ is shifted so that the normalization condition for $p$ is satisfied, as the decomposition~\eqref{C5:form}-\eqref{C5:form2} is invariant with respect to addition of constants to $V$}. Note that the generalized potential remains well-behaved even in the small noise limit. This is in contrast to the density function which becomes nearly singular and difficult to compute directly when $\epsilon$ is small.

To compute the decomposition of the force field, we parameterize the two components using neural networks. Specifically, the generalized potential $V(\vect{x})$ is approximated by
\begin{equation}\label{C5:par_V}
    V_{\theta}(\vect{x}) = \tilde{V}_{\theta}(\vect{x}) + 
\sum_{i=1}^d\rho_i (x_i -c_i)^2,
\end{equation}
where $\tilde{V}_{\theta}(\vect{x})$ is a fully-connected neural network with the activation function $\tanh$, $\rho_i$ and $c_i$ are trainable parameters, with $\rho_i > 0$.
In practice, we take $\rho_i=\log(1+\exp(\tilde{\rho}_i))$ to ensure positivity.
In Eq.~\eqref{C5:par_V}, the quadratic term is introduced so that the function $e^{-\epsilon^{-1}V_{\theta}(\vect{x})}$ is integrable in $\Rd$. Similarly, the residual component $\secondterm(\vect{x})$ is represented by a neural network $\secondterm_{\theta}(\vect{x})$. Then the force field $\vect{f}(\vect{x})$ is parameterized by
\begin{equation}\label{C5:par_f}
\vect{f}_{\theta}(\vect{x}) = -D\nabla V_{\theta}(\vect{x}) + \secondterm_{\theta}(\vect{x}).
\end{equation}

Next we introduce the loss function for training the networks under different problem settings.

\vspace{0.5cm}
\noindent{\it Fully known dynamics.}\quad 
First, consider the case when the dynamics \eqref{C5:SDE} is completely known, {\it i.e.} the force field $\vect{f}(\vect{x})$, the diffusion tensor $D$ as well as the strength of the noise $\epsilon$ are all given. \LB{This is the case considered in traditional numerical methods~\cite{sepehrian2015numerical,galan2007stochastic,torvattanabun2011numerical} and existing learning-based methods~\cite{xu2020solving,zhai2020deep}.}
To learn the parameters in $V_\theta$ and $\vect{g}_{\theta}$, we minimize the loss function
\begin{equation}\label{loss1}
\begin{aligned}
    L &= \Ldyn + \lambda \Lcondt,\\
\end{aligned}
\end{equation}
where
\begin{equation}\label{loss1b}
\begin{aligned}
    \Ldyn &= \frac{1}{d}\int_{\Rd} \lvert \vect{f}(\vect{x}) - \vect{f}_{\theta}(\vect{x}) \rvert^2 d\mu(\vect{x}),\\
\Lcondt &= \int_{\Rd}
\left\lvert
\nabla V_{\theta}(\vect{x})^T \secondterm_{\theta}(\vect{x})-\epsilon \nabla\cdot \secondterm_{\theta}(\vect{x})
\right\rvert^2 d\mu(\vect{x}),
\end{aligned}
\end{equation}
where $\mu(\vect{x})$ is a probability measure, 
$\Ldyn$ is to ensure that $\vect{f}_{\theta}$ approximates the given force field $\vect{f}$, $\Lcondt$ is to impose the constraint \eqref{C5:form2} for the decomposition of $\vect{f}$, 
and $\lambda$ is a parameter that controls the relative weights of the two terms in the loss function. 
The probability measure can be chosen at our disposal to focus on regions of interest in the dynamics.
In the numerical examples, the integrals in \eqref{loss1b} are represented as finite sums using data points sampled from the uniform distribution on a bounded domain, or a mixture of the uniformly sampled data points and those sampled from the numerical simulation of the SDE~\eqref{C5:SDE}. \LB{Note that the sampling scheme does not require a discretization mesh.}

\vspace{0.5cm}
\noindent{\it Partially known dynamics.} \quad
Next, we consider the case when the force field $\vect{f}(\vect{x})$ is unknown, but we have access to trajectory data of the deterministic dynamics corresponding to the SDE \eqref{C5:SDE}:
\begin{equation}\label{C5:det_dyn}
    \dot{\vect{x}}  = \vect{f}(\vect{x}).
\end{equation}
\LB{This is a common scenario adopted in recent works on learning dynamics from data~\cite{brunton2016discovering,yu2020onsagernet,lin2021data}.}
Furthermore we assume the diffusion tensor $D$ and the strength of the noise $\epsilon$ are given.
\LB{In the proposed method,}
we learn \LB{an interpretable dynamics with the force field} in the form of the decomposition \eqref{C5:par_f} from the trajectory data.
Specifically, we denote the observed data by 
$X=\left\{\left(X_i(t_j),X_i(t_j+\Delta t)\right): 0\leq j\leq M, 1\leq i\leq N \right\}$, which consists of $N$ trajectories $X_i$, $1\le i\le N$, each with $2M+2$ states sampled at the times $t_0,t_0+\Delta t,\dots,t_M,t_M+\Delta t$ from the dynamics \eqref{C5:det_dyn}. Here $\Delta t$ is a small time step. The $N$ trajectories start from different initial states. 
To train the force field model $\vect{f}_{\theta}$ using these data, we minimize the loss function 
\begin{equation}\label{loss2}
\begin{aligned}
L &= \Ldyn + \lambda \Lcondt,
\end{aligned}
\end{equation}
where
\begin{equation}\label{loss2b}
\begin{aligned}
     \Ldyn &= \frac{1}{N(M+1)d}\sum_{i=1}^N\sum_{j=0}^M \left\lvert
\frac{1}{\Delta t}\left(\mathcal{I}_{\Delta t}[\vect{f}_{\theta};X_i(t_j)]-X_i(t_j+\Delta t)\right)\right\rvert^2,\\
\Lcondt &=  \frac{1}{S}\sum_{k=1}^S
\left\lvert
\nabla V_{\theta}(\tilde{X}_k)^T \secondterm_{\theta}(\tilde{X}_k)-\epsilon \nabla\cdot \secondterm_{\theta}(\tilde{X}_k)
\right\rvert^2,
\end{aligned}
\end{equation}
where $\mathcal{I}_{\Delta t}[\vect{f}_{\theta};X_i(t_j)]$ is the end state obtained by performing a numerical integration of the dynamics $\dot{\vect{x}}=\vect{f}_{\theta}(\vect{x})$ by one time step $\Delta t$, starting from the state $X_i(t_j)$; $\tilde{X}=\{\tilde{X}_k\}_{k=1}^S$ is a representative subset of $X$. \LB{The data points $\tilde{X}_k$, $1\le k\le S$, in the loss $\Lcondt$ are chosen so that they are uniformly distributed in regions where the sample trajectories visited. The loss $\Lcondt$ with these representative data points can effectively 
impose the constraint $\nabla V_{\theta}(\vect{x})^T \secondterm_{\theta}(\vect{x}) - \epsilon \nabla\cdot \secondterm_{\theta}(\vect{x})=0$ in these regions. In this work, we use the algorithm proposed in Ref.~\cite{lin2021data} to sample the representative data points from the dataset $X$.}

Note that the loss function $\Ldyn$ measures the difference between the force field $\vect{f}$ and its approximation $\vect{f}_\theta$. Indeed, let $X(t)$ and $X^\theta(t)$ be the solution to the dynamics $\dot{\vect{x}}=\vect{f}(\vect{x})$ and  $\dot{\vect{x}}=\vect{f}_{\theta}(\vect{x})$, respectively, starting from the same initial state $X(0)=X^\theta(0)=X_0$. It follows that 
\begin{equation}
\begin{aligned}
    \frac{1}{\Delta t} \left|X^\theta(\Delta t)-X(\Delta t)\right| &=\frac{1}{\Delta t}\left| \int_0^{\Delta t} \left(\vect{f}_\theta(X^\theta(t))-\vect{f} (X(t)\right)dt \right|\\
    &\approx
    \left|\vect{f}_\theta(X_0) - \vect{f}(X_0)\right|.
\end{aligned}
\end{equation}
In the loss $\Ldyn$, $X^\theta(\Delta t)$ is approximated by $\mathcal{I}_{\Delta t}[\vect{f}_{\theta};X^\theta(0)]$, the solution obtained from a numerical integrator of $\dot{\vect{x}}=\vect{f}_{\theta}(\vect{x})$ starting from $X^\theta(0)$.

\section{Numerical examples}\label{C5:Numerical_Examples}

To illustrate the effectiveness of the proposed method, we apply the method to three systems with different features: a two-dimensional system with two meta-stable states, a biochemical oscillation network model, and a dynamical system in high dimensions.
In each example, we use fully connected neural networks with two hidden layers to parameterize $V$ and $\vect{g}$, and the hyperbolic tangent function ($\tanh$) as the activation function. The following three types of datasets are used in the loss function:
\begin{itemize}
\item[(i)] Data sampled from the uniform distribution on a bounded domain $\Omega$.
\item[(ii)] Data sampled from trajectories of the SDE~\eqref{C5:SDE}. 
The initial states of the trajectories are sampled from the uniform distribution on $\Omega$, and the SDE is solved using the Euler-Maruyama scheme with time step $\Delta t_1$. After the first $1000$ time steps on each trajectory, one data point is sampled for every $100$ time steps. 
\item[(iii)] Data sampled from trajectories of the deterministic dynamics~\eqref{C5:det_dyn}. The initial states of the trajectories are sampled from the uniform distribution on $\Omega$, and the dynamics is solved using the four-order Runge-Kutta method with time step $\Delta t_2$. Along each trajectory, the data points are sampled
at times $10m\Delta t_2$ and $(10m+1)\Delta t_2$, $m\geq0$. The representative data points are sampled using the algorithm in Ref.~\cite{lin2021data} with the parameter $r$.
\end{itemize}

\LB{The first and second type of datasets cover the high-probability region of interest in the dynamics. The third type is from the deterministic dynamics driven by the force field.} The domain $\Omega$, the time steps $\Delta t_1$ and $\Delta t_2$, the parameter $r$ and the network structures  are provided in Table.~\ref{C5:tab1}. We use the second-order Runge-Kutta method as the numerical integrator $\mathcal{I}$ in the loss function~\eqref{loss2b}. 
The parameter $\lambda$ in the loss function is tuned so that both the loss $\Ldyn$ and the loss $\Lcondt$ are small.
We train the neural networks using Adam optimizer~\cite{kingma2014adam} with a mini-batch of size $5000$.  The learning rate decays exponentially over the training steps.


\begin{table}[h]
	\caption{The domain $\Omega$, parameters and the network structure in the numerical examples. \LBB{Two hidden layers are used in the neural networks for all  examples.}} 
	\label{C5:tab1}
	\begin{center}
		\begin{tabular}{ cc cc c c }
			\hline\hline\vspace{-0.25cm}\\
			Example & $\Omega$ & $\Delta t_1$ & $\Delta t_2$ & $r$ &
			\makecell{ \# of nodes in each \\ hidden layer} \vspace{0.1cm}\\
			\hline\hline \vspace{-0.25cm}\\
			1 & $[-2,2]\times[-3,3]$ & $-$ & $10^{-2}$ & $0.1$ & $50$ \vspace{0.1cm}\\
			\hline \vspace{-0.25cm}\\
			2 & $[0,8]\times[0,6]$ & $10^{-3}$ & $-$ & $-$ &  $80$ \vspace{0.1cm}\\
			\hline \vspace{-0.25cm}\\
			3 & $[-2,2]^{10}$ & $-$ & $10^{-2}$ & $0.2$ &  $100$ \vspace{0.1cm}\\
			\hline\hline
		\end{tabular}
	\end{center}
\end{table}

To assess the accuracy of the learned generalized potential, we compute the relative root mean square error (rRMSE) and the relative mean absolute error (rMAE): 
\begin{equation*}
    \text{rRMSE}=\frac{\left(\int_{\mathcal{D}}\lvert V_{\theta}(\vect{x})-V(\vect{x})\rvert^2 d\vect{x}\right)^{1/2}}
    {\left(\int_{\mathcal{D}}\lvert
    V(\vect{x})\rvert^2 d\vect{x}\right)^{1/2}},\qquad
    \text{rMAE}=\frac{\int_{\mathcal{D}}\lvert V_{\theta}(\vect{x})-V(\vect{x})\rvert d\vect{x}}
    {\int_{\mathcal{D}}\lvert
    V(\vect{x})\rvert d\vect{x}},
\end{equation*}
where $V_{\theta}$ is the learned generalized potential, $V$ is the solution computed from the FP equation using the finite difference (FD) method in {\it Example 1 and 2}, and $\mathcal{D}$ is the domain $\{\vect{x}\in\Omega:V(\vect{x})\leq 20\epsilon\}$, which excludes regions of low density. To facilitate the comparison, the solutions $V_\theta$ and $V$ are shifted so that their minimum values are both 0.

In the numerical examples, the generalized potential learned using the proposed method is compared with the one computed from the corresponding FP equation using the FD method. In {\it Example 1 and 2}, the FP equation is solved on the domain $\Omega$ given in Table \ref{C5:tab1}, with the no-flux boundary condition. 
A uniform mesh with $500$ grid points in each dimension is used in the finite difference discretization. Details of the FD scheme is provided in Appendix~\ref{appendix_A}.

\subsection{\LBB{Example 1:} A two-dimensional system with two metastable states} Consider the following two-dimensional dynamical system,
\begin{equation}\label{C5:example1_SDE}
\left\{
\begin{array}{l}
\dot{x} = \dfrac{1}{5}x(1-x^2) + y(1+\sin x)+\sqrt{\dfrac{\epsilon}{5}}\xi_1, \vspace{0.15cm} \\
\dot{y} = -y+2x(1-x^2)(1+\sin x)+\sqrt{2\epsilon}\ \xi_2,
\end{array} \right.
\end{equation}
where the state of the system is $\vect{x}=(x,y)^T$, $\bm\xi=(\xi_1,\xi_2)^T$ is a two-dimensional white noise, and the diffusion tensor $D=\diag(0.1,1)$.
The corresponding deterministic dynamics ($\epsilon=0$) has two stable stationary points at $\vect{x}_a=(-1,0)^T$ and $\vect{x}_b=(1,0)^T$ and one unstable stationary point at $\vect{x}_c=(0,0)^T$. 

First we assume the dynamics in \eqref{C5:example1_SDE} is completely known to us. In this case, we use the loss function \eqref{loss1}-\eqref{loss1b} to train the neural networks for $\vect{f}_\theta$. 
The integrals in the loss function are represented as finite sums using $10^4$ data points sampled from the uniform distribution on $\Omega=[-2,2]\times [-3,3]$ (dataset (i)). 
Using these data, we train the 
\LB{model $\vect{f}_{\theta}$} at $\epsilon=0.2$, $\epsilon=0.1$ and $\epsilon=0.05$, respectively. 

\begin{figure}[t!]
	\centering
 	\includegraphics[width=\linewidth]{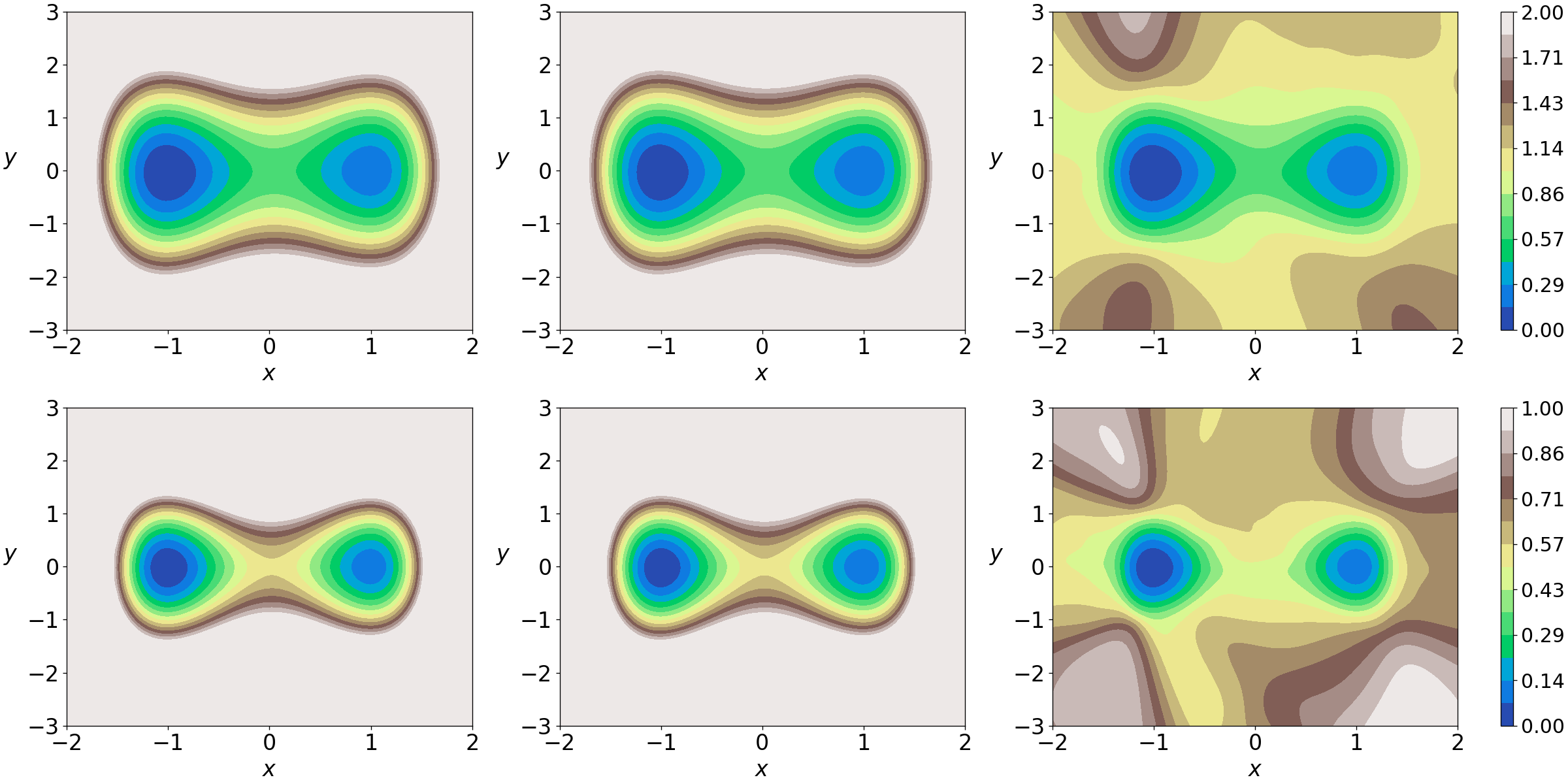}
 	\caption{(Example 1) Contour plots of the potential
 	$V(\vect{x})=-\epsilon\log p(\vect{x})$, where $p(\vect{x})$ is the finite difference solution of the FP equation (left), 
 	$V_{\theta}(\vect{x})$ learned using the loss function \eqref{loss1}-\eqref{loss1b} (middle) and the potential $V'(\vect{x})=-\epsilon\log p_{\theta}(\vect{x})$, where $p_\theta(\vect{x})$ is learned using the loss function \eqref{loss_r2} (right). The noise is $\epsilon=0.1$ (top) and $\epsilon=0.05$ (bottom).
	}
	\label{C5:fig3}
\end{figure}

The learned generalized potentials $V_\theta$ for $\epsilon=0.1$ and $0.05$ are shown in Fig.~\ref{C5:fig3}. Also shown in the figure are the solution $V=-\epsilon\log p$ obtained by solving the FP equation for $p$ using the finite difference method (the FD solution) and the solution $V'=-\epsilon\log p_\theta$, where $p_\theta$ is computed by minimizing the loss function \eqref{loss_r2}. In the loss function \eqref{loss_r2}, we use the FD solution as the estimator $\tilde{p}$, and the invariant density function $p_\theta$ is trained directly. This method gives less accurate solution as compared to the proposed method for learning the generalized potential, especially when the temperature is low and the density is narrowly peaked at the metastable states, as can be seen from the numerical results. Details of training the invariant density function using the loss function \eqref{loss_r2} are provided in Appendix~\ref{appendix_B}. 
A quantitative assessment of the numerical solutions is provided in Table~\ref{C5:tab2}, where we report the root mean square error and the root mean absolute error of the learned potentials. For each value of $\epsilon$, the FD solution is used as the reference solution to compute the errors of the learned potential. \LB{Each error in the table is computed from $10$ independent runs including sampling the data and training the networks.} The advantage of parameterizing the potential over the method of parameterizing the invariant density is also evident from the results shown in the table.

\begin{table}[h]
	\caption{Example 1: \LB{The root mean square error and root mean absolute error of the potentials $V_{\theta}$ and $V'=-\epsilon\log p_{\theta}$, where $V_\theta$ and $p_\theta$ are learned using the loss function \eqref{loss1}-\eqref{loss1b} and \eqref{loss_r2}, respectively. The statistics (mean$\pm$ deviation) is based on $10$ independent runs.} The parameter $\lambda=1$.}
	\label{C5:tab2}
	\begin{center}
		\begin{tabular}{ cccc c }
			\hline\hline\vspace{-0.25cm}\\
			$\epsilon$  & rRMSE of $V_{\theta}$ & rMAE of $V_{\theta}$ & rRMSE of $V'$ & rMAE of $V'$  \\
			\hline\hline \vspace{-0.25cm}\\
			$0.2$  & $0.0073\pm 0.0029$ & $0.0072\pm 0.0033$ & $0.3960\pm 0.0338$ & $0.2533 \pm 0.0237$
			\\
			\hline \vspace{-0.25cm}\\
			$0.1$  & $0.0107\pm 0.0043$ & $0.0102\pm 0.0040$ & $0.3789\pm 0.0335$ & $0.2460\pm 0.0228$
			\\
			\hline \vspace{-0.25cm}\\
			$0.05$  & $0.0194\pm 0.0084$ & $0.0193\pm 0.0090$ & $0.3903\pm 0.0315$ & $0.2790\pm 0.0270$
			\\
			\hline\hline
		\end{tabular}
	\end{center}
\end{table}

\begin{figure}[t!]
	\centering
 	\includegraphics[width=\linewidth]{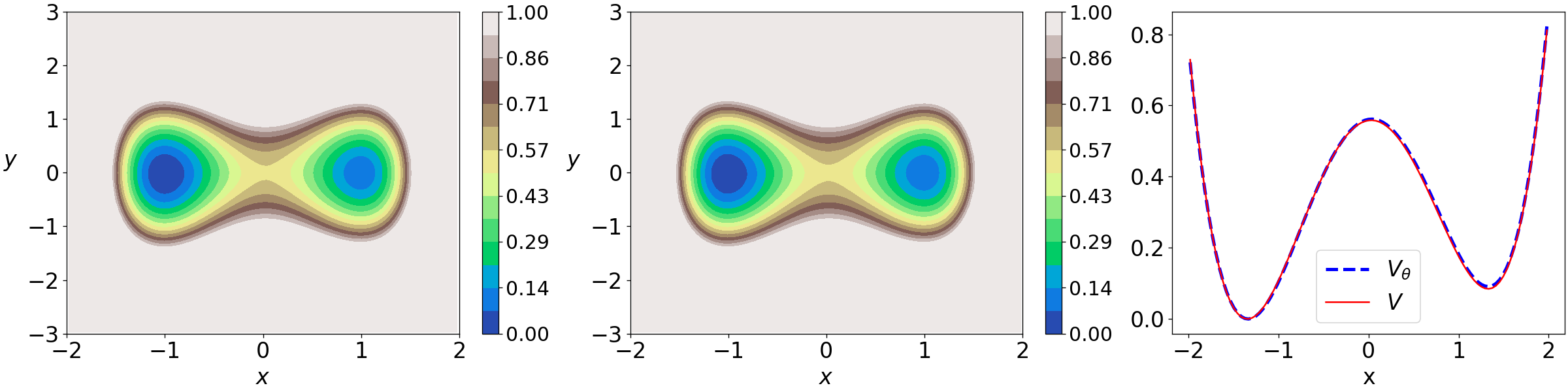}
	\caption{(Example 1) Contour plots of the potential $V_{\theta}$ learned using the loss function \eqref{loss2}-\eqref{loss2b} (left), the potential $V$ computed using the FD method (middle) and plots of $V_{\theta}(x, y=0)$ and $V(x, y=0)$ (right). The noise $\epsilon$ is 0.05. The parameter $\lambda$ is $0.1$. 
	}
	\label{C5:fig4}
\end{figure}

Next we assume that the force field in \eqref{C5:example1_SDE} is unknown to us. In this case, we compute the neural network model for the force field, $\vect{f}_\theta$, in the form of the decomposition \eqref{C5:par_f} by minimizing the loss function \eqref{loss2}-\eqref{loss2b}. 
The dataset contains $10^5$ data points sampled from $500$ trajectories of the deterministic dynamics (dataset (iii)).
Using these data, we train the neural network model $\vect{f}_{\theta}$ at $\epsilon=0.05$, \LBB{$\epsilon=0.1$ and $\epsilon=0.2$, respectively}.
The learned potential $V_{\theta}(\vect{x})$ \LBB{for $\epsilon=0.05$} is shown in Fig.~\ref{C5:fig4}, together with the finite difference solution for the purpose of comparison.  The root mean square error and the root mean absolute error of $V_{\theta}$ are $0.0095$ and $0.0084$, respectively.

\begin{figure}[t!]
	\centering
 	\includegraphics[width=.8\linewidth]{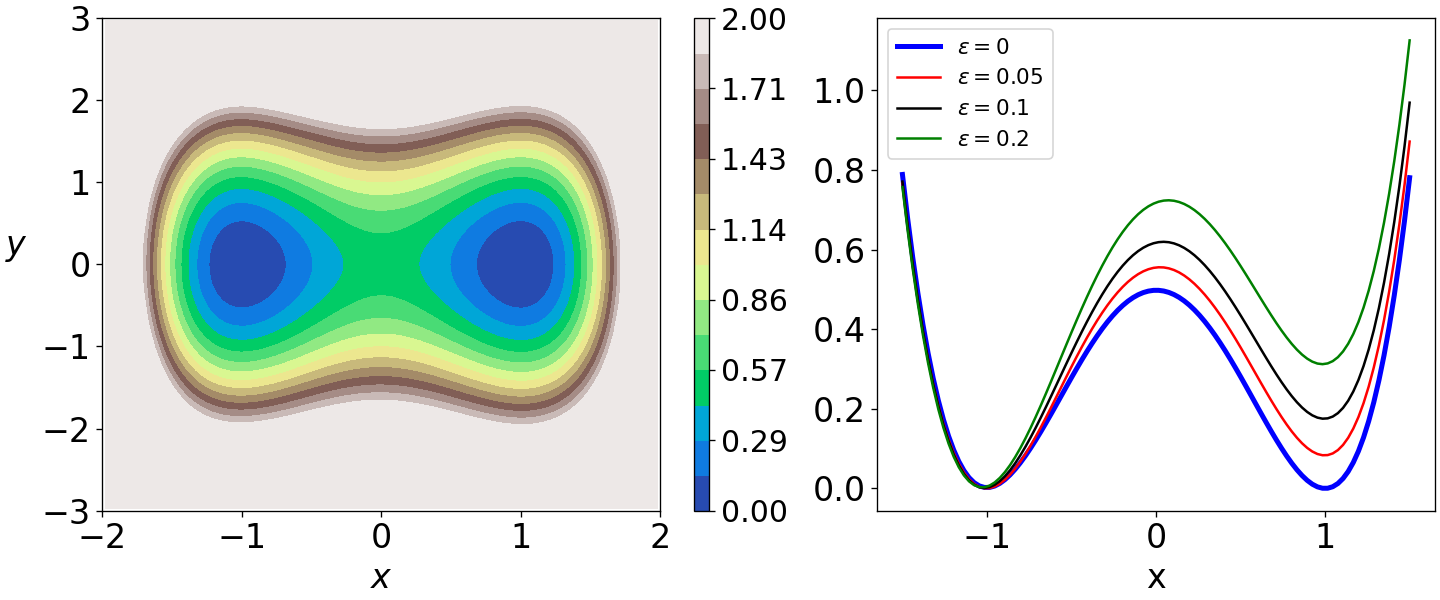}
 	\caption{\LBB{(Example 1) Contour plot of the quasipotential $U_{\theta}(\vect{x})$ computed using the method in Ref.~\cite{lin2021data} (left) and plots of $U_{\theta}(\vect{x})$ ($\epsilon=0$) and the learned generalized potentials $V_{\theta}(\vect{x})$ for different values of $\epsilon$ ($\epsilon=0.05,0.1,0.2$) along the line $y=0$ (right).}
	}
	\label{fig1}
\end{figure}

\LBB{With the sampled trajectory data, we also apply the method in Ref.~\cite{lin2021data} to compute the respective quasipotentials associated with each attractor of the system. 
The learned quasipotential $U_{\theta}(\vect{x})$ is shown in the left panel of Fig.~\ref{fig1}. In particular, we plot the quasipotential $U_{\theta}(\vect{x})$ and the generalized potential $V_{\theta}(\vect{x})$ for $\epsilon=0.05,0.1,0.2$ along the line $y=0$ in the right panel of Fig.~\ref{fig1}. From the figure, we can see that as the noise tends to zero, the generalized potential converges to the global quasipotential. Also, the numerical results reveal that the finite noise in the system~\eqref{C5:example1_SDE} has a significant entropic effect on the landscape of the equilibrium distribution: the left well of the system concentrates more equilibrium probabilities than the right one, as the magnitude of the noise increases.
}

\subsection{\LBB{Example 2:} A biochemical oscillation network model}

To show the effectiveness of the proposed method in systems with other features, we evaluate our method on a system whose potential landscape has a limit-cycle shape when the temperature is low. We consider a biochemical oscillation network of cell cycles~\cite{wang2008potential}. The network consists of two cyclins: CLN/CDC28 and CLB/CDC28. Let $x$ and $y$ denote the average concentration of the two cyclins, respectively. The variation rates of the concentrations are described by the SDEs
\begin{equation}\label{C5:example2_SDE}
\left\{
\begin{array}{l}
\dot{x} = 100\left(\dfrac{\alpha^2+x^2}{1+x^2}\dfrac{1}{1+y}-ax\right)+\sqrt{2\epsilon}\ \xi_1, \vspace{0.15cm} \\
\dot{y} = \dfrac{100}{\tau_0}\left(b-\dfrac{y}{1+cx^2}\right)+\sqrt{2\epsilon} \ \xi_2,
\end{array} \right.
\end{equation}
where the state of the system is $\vect{x}=(x,y)^T$, $\bm\xi=(\xi_1,\xi_2)^T$ is a two-dimensional white noise, and the diffusion tensor $D=\diag(1,1)$. The parameters in the equations are taken as $\alpha=0.1$, $a=0.1$, $\tau_0=5$, $b=0.1$ and $c=100$. 

\begin{figure}[t!]
	\centering
 	\includegraphics[width=\linewidth]{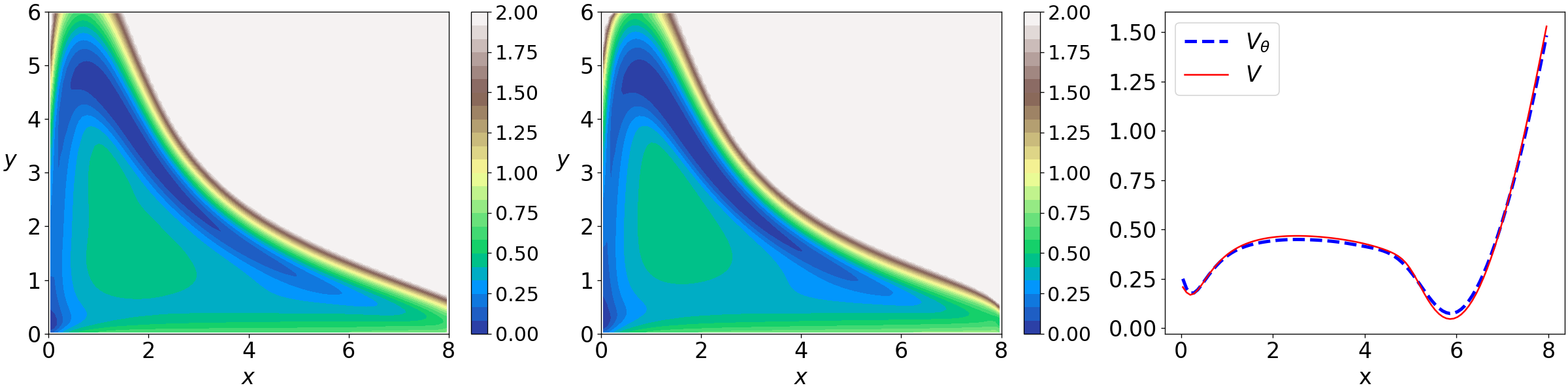}
	\caption{(Example 2) Contour plots of the potential $V_{\theta}$ learned using the loss function \eqref{loss1}-\eqref{loss1b} (left), the potential $V$ computed using FD method (middle) and plots of $V_{\theta}(x, y=2)$ and $V(x, y=2)$ (right). The noise $\epsilon$ is 0.1.
	}
	\label{C5:fig5}
\end{figure}

We assume the dynamics in \eqref{C5:example2_SDE} 
is completely known to us. We use the loss function \eqref{loss1}-\eqref{loss1b} to train the neural network model $\vect{f}_\theta$ in the form of the decomposition \eqref{C5:par_f}. The integrals in the loss function are represented as finite sums using a mixture of 
$2\times 10^3$ data points sampled from the uniform distribution on $\Omega = [0,8]\times [0,6]$ (dataset (i)) and $8\times 10^3$ 
data points sampled from trajectories of the SDEs~\eqref{C5:example2_SDE} at the temperature $\epsilon=0.1$ (dataset (ii)). \LB{To focus on the region of interest, only the data points inside the region $\{\vect{x}:\inf_{\vect{y}\in\Omega}\lVert\vect{x}-\vect{y}\rVert_2\leq1\}$ are kept in sampling of the SDE data}.
Using these data, we train the force field model $\vect{f}_{\theta}$ at $\epsilon=0.1$. 
The numerical solution for the potential $V_{\theta}(x,y)$ is shown in Fig.~\ref{C5:fig5}. Also shown in the figure is the finite difference solution $V(x,y)$. The two solutions agree very well.

We also conducted the computation for \LB{$\epsilon=0.2$ and $\epsilon=0.3$, respectively.}
The errors of $V_{\theta}(x,y)$ for the different values of $\epsilon$ are reported in Table~\ref{C5:tab3}.
These results show that our method is effective in capturing different types of potential landscapes.

\begin{table}[h]
	\caption{Example 2: The errors of the learned potential $V_{\theta}$ for different values of $\epsilon$. The statistics (mean$\pm$ deviation) is based on $10$ independent runs. The parameter $\lambda$ is 0.3.}
	\label{C5:tab3}
	\begin{center}
		\begin{tabular}{ cccc c }
			\hline\hline\vspace{-0.25cm}\\
			$\epsilon$   & rRMSE of $V_{\theta}$ & rMAE for $V_{\theta}$\\
			\hline\hline \vspace{-0.25cm}\\
			$0.3$  & $0.1505\pm 0.0117$ & $0.1037\pm 0.0072$ 
			\\
			\hline \vspace{-0.25cm}\\
			$0.2$  & $0.1275\pm 0.0320$ & $0.0808\pm 0.0154$ 
			\\
			\hline \vspace{-0.25cm}\\
			$0.1$  & $0.0897\pm 0.0283$ & $0.0663\pm 0.0154$ 
			\\
			\hline\hline
		\end{tabular}
	\end{center}
\end{table}

\newcommand{\Amatrix}{B}
\subsection{\LBB{Example 3:} A ten-dimensional system}
To show the effectiveness of the proposed method in high-dimensional systems where traditional numerical methods are not directly applicable, we consider a dynamical system in ten-dimensional space $\mathbb{R}^{10}$.
We choose a synthetic example with an explicitly known reduction to a lower dimensional system, on which the invariant
distribution can be computed accurately.
This is important for validating the proposed method in high dimensions, since classical
numerical methods such as finite difference and finite element methods cannot be directly applied to obtain reference solutions for general high dimensional systems.
Concretely, we consider the ten-dimensional system
\begin{equation}\label{C5:Example3_SDE_y}
\dot{\vect{x}}=
\Amatrix \vect{h}(\Amatrix^{-1} \vect{x}) + \sqrt{2\epsilon}\Amatrix \bm\xi, \quad t>0,
\end{equation}
where $\vect{h}(\vect{y})=(h_1(\vect{y}),\dots,h_{10}(\vect{y}))^T$ is a vector field with 
\begin{equation*}
\begin{aligned}
h_{2k-1}(\vect{y})&= v_1(y_{2k-1},y_{2k}):= -y_{2k-1}+y_{2k}(1+\sin y_{2k-1}),\\
h_{2k}(\vect{y})  &= v_2(y_{2k-1},y_{2k}):=-y_{2k}-y_{2k-1}(1+\sin y_{2k-1}), \quad 1\leq k\leq 5,
\end{aligned}
\end{equation*}
$\Amatrix=[b_{i,j}]$ is a $10{\times}10$ matrix given by
\begin{equation*}
b_{i,j} = \left\{
\begin{array}{cl}
0.8,  & \text{for}\ i=j=2k-1, 1\le k\le 5, \\
1.25, & \text{for}\ i=j=2k, 1\le k\le 5, \\
-0.5, & \text{for}\ j=i+1, 1\le i\le 9\\
0,    & \text{otherwise},
\end{array} \right.
\end{equation*}
and $\bm\xi=(\xi_1,\dots,\xi_{10})^T$ is a ten-dimensional white noise.
This system is obtained by coupling five independent two-dimensional systems of the same form:
\begin{equation}\label{exmaple3_2d}
\left\{
\begin{array}{l}
\dot{y}_{2k-1} =v_1(y_{2k-1}, y_{2k}) + \sqrt{2\epsilon}\ \xi_{2k-1},\\
\dot{y}_{2k} \ \ \ \ = v_2(y_{2k-1}, y_{2k}) +\sqrt{2\epsilon}\ \xi_{2k},\qquad 1\leq k\leq 5,
\end{array} 
\right.
\end{equation}
and $\vect{x}=B\vect{y}$. As shown in Appendix~\ref{appendix_C}, the generalized potential of the system~\eqref{C5:Example3_SDE_y} is given by
\begin{equation}\label{exmaple3_V}
    V(\vect{x})=V_0(y_1,y_2)+\dots+V_0(y_9,y_{10}),
\end{equation}
where $(y_1,\dots,y_{10})^T=\Amatrix^{-1}\vect{x}$ and $V_0$ is the generalized potential of the two-dimensional system~\eqref{exmaple3_2d}. Thus a reference solution of $V$ can be obtained by solving the FP equation associated with the two-dimensional system using the finite difference method.

\begin{figure}[t!]
	\centering
	\includegraphics[width=\linewidth]{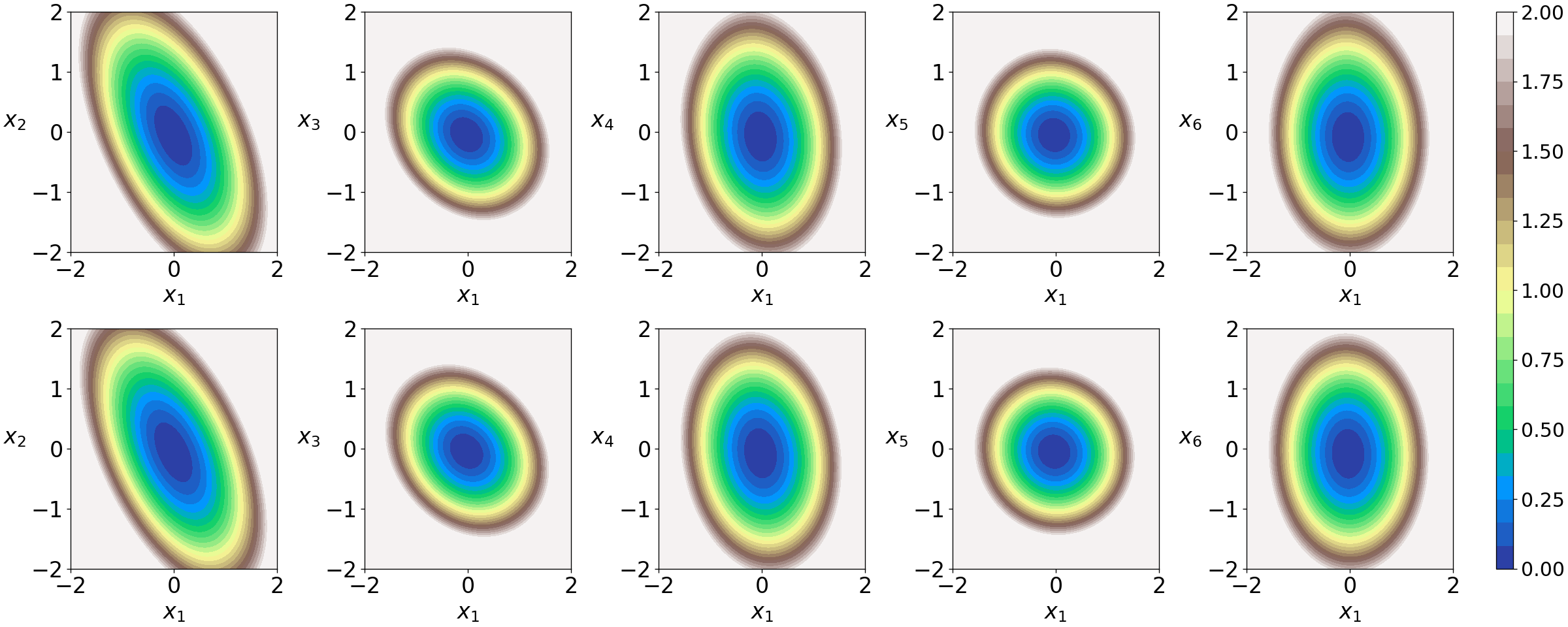}
	\caption{(Example 3) Cross sections of the potential $V_{\theta}$ learned using the loss function \eqref{loss2}-\eqref{loss2b} (top) and the reference solution $V$ computed using Eq.~\eqref{exmaple3_V} (bottom). 
	The cross sections are taken at $(x_1, x_i)\in \mathbb{R}^2$, $2\le i\le 6$, with the other coordinates being 0. The noise $\epsilon=0.1$ and $\lambda=0.1$. 
	}
	\label{C5:fig7}
\end{figure}

We assume the force field in Eq.~\eqref{C5:Example3_SDE_y} is unknown to us. In this case, we learn the neural network model for the force field, $\vect{f}_{\theta}$, in the form of the decomposition~\eqref{C5:par_f} by minimizing the loss function~\eqref{loss2}-\eqref{loss2b}. 
The dataset contains $10^6$ data points sampled from $10^4$ trajectories of the deterministic dynamics (dataset (iii)).
Using these data, we train the neural network model $\vect{f}_{\theta}$ at $\epsilon=0.1$.
As the learned potential $V_\theta(\vect{x})$ is in the ten-dimensional space, we plot a number of its cross sections in Fig.~\ref{C5:fig7}. Also shown in the figure are cross sections of the reference solution $V$.
\LB{It can be observed from the results that the two solutions agree well in the five cross sections in both high-probability and low-probability regions for this high-dimensional system.}

\section{Conclusion}\label{C5:Conclusion}
In this paper, we developed a machine learning method to compute the invariant distribution of randomly perturbed dynamical systems modeled by SDEs. We considered two scenarios: in the first one, the force field is given;  while in the second one, the force field is unknown but we have access to the data of the deterministic dynamics. In each case, we proposed an appropriate loss function. The method learns the force field in the form of a decomposition as suggested by the FP equation by minimizing the loss function. The two components of the decomposition are parameterized by neural networks. The potential component of the decomposition
gives the generalized potential for the invariant distribution. 

The proposed method was shown to be effective in various systems with different features, including a dynamical system with two meta-stable states, a biological network model, and a system in high dimensions. In all these examples, the numerical results agreed well with the respective reference solutions. The advantage of parameterizing the generalized potential rather than the invariant density function directly was also demonstrated in one of the examples. \LB{Furthermore,} the method is data-driven in the sense that it does not require any prior knowledge of the force field other than the data of the dynamics. 
The method enables us to study the equilibrium properties of practical dynamical systems in high dimensions \LB{at low temperatures}. 

\LBB{In the current work, we considered systems with known noise covariance structure. Also, the data was sampled from the deterministic dynamics. In an ongoing work, we extend this method to learn the invariant distribution together with the structure of the noise from noisy data.}

\appendix
\section{Solving the Fokker-Planck equation using the finite difference method}\label{appendix_A}
Consider the two-dimensional Fokker-Planck equation, 
\begin{equation}\label{eq1}
	\nabla\cdot \mathbf{J} = 0, \quad (x,y)\in \Omega
\end{equation}
where $\Omega=[a, a+L_x]\times[b, b+L_y]$, and the probability flux $\mathbf{J}(x,y)=[J_1,J_2]^T$ is given by
\begin{equation}\label{flux}
	\begin{aligned}
		J_1&=-f_1(x,y)p(x,y) + \epsilon_1 \partial_{x}p(x,y),\\
		J_2&=-f_2(x,y)p(x,y) + \epsilon_2 \partial_{y}p(x,y).
	\end{aligned}
\end{equation}
Equation~\eqref{eq1} is supplemented with the no-flux boundary condition 
\begin{equation}\label{eq2}
	\mathbf{J}(x,y)\cdot \mathbf{n}(x,y) =0,\quad (x,y)\in\partial\Omega
\end{equation}
where $\mathbf{n}$ is the outward normal of the boundary of $\Omega$, and the normalization condition
\begin{equation} \label{norm}
	\int_{\Omega}p(x,y)dxdy=1.
\end{equation}
We discretize the domain $\Omega$ using a uniform mesh with grid points $(x_i,y_j)$, $0\leq i\leq N_x$, $0\leq j\leq N_y$, where $x_i=ih_x$, $y_j=jh_y$, $h_x=L_x/N_x$, $h_y=L_y/N_y$.
We denote the mid-points of the mesh (i.e. cell centers) by $(x_{i-1/2}, y_{j-1/2})$, where $x_{i-1/2}=x_i-h_x/2$, $1\leq i\leq N_x$, and $y_{j-1/2}=y_j-h_y/2$, $1\leq j\leq N_y$. 
The solution for $p$ is computed at the cell centers.

The flux $J_1(x,y)$ at the point $(x_i,y_{j-1/2})$, $1\leq i\leq N_x-1,\ 1\leq j\leq N_y$ is approximated by
\begin{equation}\label{flux1_d}
	\begin{aligned}
		J_1^{i,j-1/2} = &-f_1(x_i,y_{j-1/2})\cdot\frac{1}{2}\left(p_{i-1/2,j-1/2}+p_{i+1/2,j-1/2}\right)\\
		&+ \frac{\epsilon_1}{h_x}\left(p_{i+1/2,j-1/2}-p_{i-1/2,j-1/2}\right),
	\end{aligned}
\end{equation}
where $p_{i-1/2,j-1/2}$ denotes the approximate solution for $p(x,y)$ at the cell center $(x_i-h_x/2,y_j-h_y/2)$.
Similarly, the flux $J_2(x,y)$ at the point $(x_{i-1/2},y_j)$, $1\leq i\leq N_x,\ 1\leq j\leq N_y-1$ is approximated by
\begin{equation}\label{flux2_d}
	\begin{aligned}
		J_2^{i-1/2,j} = &-f_2(x_{i-1/2},y_j)\cdot\frac{1}{2}\left(p_{i-1/2,j-1/2}+p_{i-1/2,j+1/2}\right)\\
		&+ \frac{\epsilon_2}{h_y}\left(p_{i-1/2,j+1/2}-p_{i-1/2,j-1/2}\right).
	\end{aligned}
\end{equation}
In the Fokker-Planck equation~\eqref{eq1}, we approximate the derivatives using the centered difference. This yields 
\begin{equation}\label{eq2_d}
	\begin{aligned}
		\frac{1}{h_x}\left(J_1^{i,j-1/2}-J_1^{i-1,j-1/2}\right) &+ 
		\frac{1}{h_y}\left(J_2^{i-1/2,j}-J_2^{i-1/2,j-1}\right) = 0,\\
		&1\leq i\leq N_x,\ 1\leq j\leq N_y.
	\end{aligned}
\end{equation}
The no-flux boundary condition gives
\begin{equation}\label{bd2_d}
	\begin{aligned}
		J_1^{0,j-1/2}&=J_1^{N_x,j-1/2}=0,\quad 1\leq j\leq N_y,\\
		J_2^{i-1/2,0}&=J_2^{i-1/2,N_y}=0,\quad 1\leq i\leq N_x.
	\end{aligned}
\end{equation}
Equations \eqref{eq2_d}-\eqref{bd2_d} form a linear system 
\begin{equation}\label{LS4_d}
	A\mathbf{p}=0,
\end{equation}
where $A\in \mathbb{R}^{N\times N}$, $N=N_x\cdot N_y$, and $\vect{p}\in \mathbb{R}^N$ 
is the vector formed by $(p_{i-1/2, j-1/2})$, $1\le i\le N_x$, $1\le j\le N_y$. 
The normalization condition~\eqref{norm} is approximated by
\begin{equation}\label{norm2_d}
	h_xh_y\sum_{p\in \mathbf{p}}p = 1.
\end{equation}
Equations \eqref{LS4_d}-\eqref{norm2_d} determine the unique solution for $\vect{p}$. We solve these equations to obtain an approximate solution for the invariant density function.

\section{Computing the generalized potential using the loss function~\ref{loss_r2}}\label{appendix_B}
In {\it Example 1}, for the purpose of comparison, we implemented the method proposed in Ref.~\cite{zhai2020deep} with the loss function~\ref{loss_r2}. 
The invariant distribution is parameterized by $p_{\theta}(x,y)=\log(1+\exp(\tilde{p}_{\theta}(x,y)))$,
where $\tilde{p}_{\theta}$ is a vanilla neural network with two hidden layers and the activation $\tanh$. Each hidden layer has $50$ nodes. The positive function acting on the output of the network is introduced to guarantee the positivity of the density function. In the loss function~\ref{loss_r2}, we replace the Monte Carlo estimator $\tilde{p}(x,y)$ with the finite difference solution $p(x,y)$ and take
\begin{equation}\label{loss_r2_emp}
	L = \frac{1}{N}\sum_{i=1}^{N}\lvert \mathcal{N}p_{\theta}(\vect{x}_i)\rvert^2 + 
	\frac{1}{M}\sum_{i=1}^{M}\lvert p_{\theta}(\vect{y}_i)-p(\vect{y}_i)\rvert^2,
\end{equation}
where $\{\vect{x}_i\}_{i=1}^{N}$, $N=10^4$ and $\{\vect{y}_i\}_{i=1}^{M}$, $M=500$ are data points sampled from the uniform distribution on $\Omega = [-2,2]\times[-3,3]$. The training problem is solved by the ``double shuffling'' method used in Ref.~\cite{zhai2020deep}, which alternatively performs gradient-descent steps for minimizing the two separate terms in the loss function~\eqref{loss_r2_emp}. The contour plots of $V'=-\epsilon\log p_{\theta}$ on $\Omega$ at $\epsilon=0.1$ and $\epsilon=0.05$ are shown in the right panel of Fig.~\ref{C5:fig3} in the paper.

\section{Proof of equation~\ref{exmaple3_V}}\label{appendix_C} 
Consider the following  dynamical system in the $d$-dimensional space,
\begin{equation}
	\dot{\vect{y}}=
	\vect{h}(\vect{y}) + \sqrt{2\epsilon}\ \bm\xi, \quad t>0 \label{system2}
\end{equation}
where $\xi$ is a $d$-dimensional white noise. Let $B$ be a $d\times d$ constant matrix with $\det B=1$, and $\vect{x}(t) = B\vect{y}(t)$. Then the process $\vect{x}(t)$ satisfies the equation
\begin{equation}
	\dot{\vect{x}}=
	\Amatrix \vect{h}(\Amatrix^{-1} \vect{x}) + \sqrt{2\epsilon}\Amatrix \bm\xi,\quad t>0. \label{system1}
\end{equation}
The Fokker-Planck equation associated with the system~\eqref{system2} and the system 
\eqref{system1} is respectively given by
\begin{alignat}{2}
	-\nabla_{\vect{y}}\cdot(\vect{h}(\vect{y})p_{\vect{y}}(\vect{y})) + \epsilon\nabla_{\vect{y}}\cdot(\nabla_{\vect{y}} p_{\vect{y}}(\vect{y}))&=0, \quad \vect{y}\in \mathbb{R}^d, \label{FP_y} \\
	-\nabla_{\vect{x}}\cdot (\Amatrix \vect{h}(\Amatrix^{-1} \vect{x})p_{\vect{x}}(\vect{x})) + \epsilon\nabla_{\vect{x}}\cdot(\Amatrix \Amatrix^T\nabla_{\vect{x}} p_{\vect{x}}(\vect{x}))&=0,\quad  \vect{x}\in \mathbb{R}^d \label{FP_x}.
\end{alignat}
It is straightforward to verify that 
\begin{equation}\label{relation1}
	p_{\vect{x}}(\vect{x})=p_{\vect{y}}(B^{-1}\vect{x}).
\end{equation}

In {\it Example 3}, the system~\eqref{system2} is composed of five independent two-dimensional systems of same form, therefore the probability density $p_\vect{y}$ is given by
\begin{equation}\label{relation2}
	p_{\vect{y}}(\vect{y}) = \prod_{k=1}^5 p_0(y_{2k-1},y_{2k}), 
\end{equation}
where $p_0$ is the invariant distribution of the  two-dimensional system. Let $V_0=-\epsilon\log p_0$. Then from Eqns.~\eqref{relation1}-\eqref{relation2}, the generalized potential $V_{\vect{x}}=-\epsilon\log p_{\vect{x}}$ is given by
\begin{equation}
	V_{\vect{x}}(\vect{x}) = \sum_{k=1}^5 V_0(y_{2k-1},y_{2k}),
\end{equation}
where $(y_1,\dots,y_{10})^T=\Amatrix^{-1}\vect{x}$.


\LBB{
\section*{Acknowledgements}
The work of B. Lin and W. Ren were partially supported by A*STAR under its AME Programmatic programme: Explainable Physics-based AI for Engineering Modelling \& Design (ePAI) [Award No. A20H5b0142].}
Q. Li is supported by the National Research Foundation of Singapore, under the NRF Fellowship NRFF13-2021-0106.


\bibliographystyle{siamplain}
\bibliography{references}

\end{document}